\documentclass[fleqn,twoside]{article}
\usepackage{espcrc2}

\usepackage{graphicx}
\usepackage[figuresright]{rotating}

\newcommand{\sleq} {\raisebox{-.6ex}{${\textstyle\stackrel{<}{\sim}}$}}

\title{Numerical test of Polyakov loop models in high temperature SU(2)}

\author{Roberto Fiore\address[CS]{Dipartimento di Fisica, Universit\`a della 
Calabria \\ 
\& Istituto Nazionale di Fisica Nucleare, Gruppo Collegato di Cosenza, Italy}, 
Pietro Giudice\addressmark[CS] and Alessandro Papa\addressmark[CS]}
       
\begin{document}

\begin{abstract}
We study the compatibility of effective mean-field models of the Polyakov loop 
for the deconfined phase of SU(N) pure gauge theories with lattice data 
obtained for the case of SU(2), in the temperature range $T_c\div 4.8 T_c$. 
\end{abstract}

\maketitle

\section{INTRODUCTION}

It has been suggested in several papers (see Ref.~\cite{Pis02})
that the deconfined phase of SU(N) pure gauge theories could be
described by an effective {\em mean-field} theory of the Polyakov loop, 
possessing global Z(N) invariance.
Through this effective theory, a relation can be established between the 
pressure of the gluon gas and the Polyakov loop. 
If the phase transition is second order as for SU(2) or ``weakly'' first 
order as for SU(3), this effective theory can be written near the transition
in terms of the first few powers of the Polyakov loop and 
of its complex conjugate. In this case, the relation between pressure and 
Polyakov loop becomes very simple and its compatibility with lattice data 
can be easily tested. 

In this study, we have considered the case of SU(2) pure
gauge theory on a 16$^3\times4$ lattice with the standard Wilson action,
in the temperature range $T_c \div 4.80\,T_c$. Although lattice effects 
are large for the Wilson action with $N_\tau=4$ sites in the time direction, 
the shape of the behavior of pressure and Polyakov loop with the temperature
should not be different from the cases of larger values of $N_\tau$, as seen
in SU(3)~\cite{Kar02}.

\section{LATTICE DETERMINATIONS}

\begin{figure}[htb]
\includegraphics*[scale=0.4]{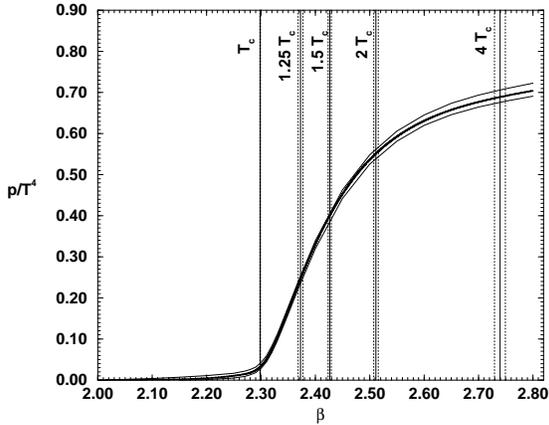}
\vspace{-1.7cm}\caption{The three solid curves represent $p/T^4$ and its
uncertainty; the vertical lines represent the critical couplings
on lattices with $N_\tau$=4, 5, 6, 8, 16~\cite{FHK93}.}
\label{pressure}\vspace{-0.5cm}
\end{figure}

The pressure of the gluon gas is given by
\begin{equation}
\frac{p}{T^4}=-\left. \frac{f}{T^4}\right|_{\beta_0}^\beta = 
N_\tau^4\int_{\beta_0}^\beta d\beta' [\langle S_0\rangle 
-\langle S_T\rangle ]\;,
\end{equation}
where $S_0$ ($S_T$) is the action density at zero (non-zero) temperature, 
$\beta_0$ is an arbitrarily chosen value, small enough that the integrand 
function at this point has become zero. Monte Carlo simulations were
performed on $16^4$ lattices for zero-temperature (typical statistics 30K), 
and on $16^3\times 4$ lattices for non-zero temperature 
(typical statistics 80K). Numerical results for $N_\tau^4 [\langle S_0\rangle 
-\langle S_T\rangle ]$ were interpolated by cubic splines before
the numerical integration which led to the pressure (Fig.~\ref{pressure}). 
As an estimate of the uncertainty for the pressure, we calculated also the
integral by interpolating the data for $N_\tau^4 [\langle S_0\rangle 
-\langle S_T\rangle ]$ with the broken line connecting the 1$\sigma$ 
upper (lower) bound of each determination. 
The correspondence between $\beta$ and the temperature has been established 
using the interpolating ansatz of Ref.~\cite{EKR95}, which makes use of the  
known~\cite{FHK93} critical couplings on lattices with $N_\tau$=4, 5, 6, 8, 
16.

\begin{figure}[htb]
\includegraphics*[scale=0.4]{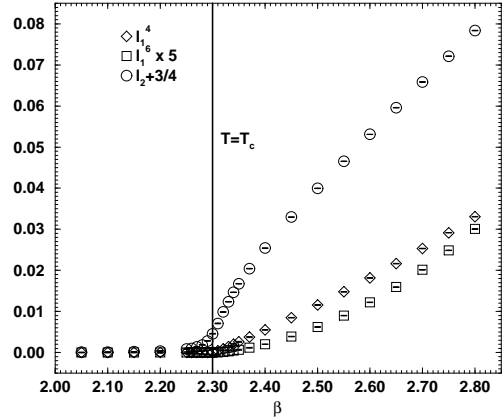}
\vspace{-1.7cm}\caption{$l_1^4$, $l_1^6$ and $l_2$ {\it vs} $\beta$ on 
a $16^3\times 4$ lattice.}
\label{loops}\vspace{-0.7cm}
\end{figure}

We considered both the charge-1 and charge-2 Polyakov loops, given
respectively by $l_1=\frac{1}{2} \langle \mbox{Tr} L(\vec{x})\rangle$
and $l_2=\frac{1}{2} \langle \mbox{Tr} L(\vec{x})^2\rangle 
- \left[\frac{1}{2} \langle \mbox{Tr} L(\vec{x})\rangle\right]^2 $,
with $L(\vec x)=\prod_{n_4=1}^{N_\tau} U_4(\vec x,n_4) $. 
We observe that $l_2$ is Z(2)-invariant and is connected to the 
Polyakov loop in the adjoint color representation by 
$l_{\mbox{\footnotesize adj}} = 1 + 4 l_2/3$.
In Fig.~\ref{loops} we show the behavior of $l_1^4$, $l_1^6$ and $l_2$ 
with $\beta$. We observe that $l_1^4$ goes linear in the region 
$ 2.30\,\sleq\,\beta \,\sleq\,2.37$ (corresponding to $T_c \,\sleq\, T 
\,\sleq\, 1.27 \,T_c$) and in the region $2.45\,\sleq\,\beta\,\sleq\,2.80$ 
(corresponding to $1.65 \,T_c \,\sleq\, T \,\sleq\, 4.80 \,T_c$). Moreover,
$l_2$ goes to $-3/4$ in the confined phase, thus implying 
$l_{\mbox{\footnotesize adj}}\to 0$ in that phase (for details on the 
behavior of $l_{\mbox{\footnotesize adj}}$ across the transition, 
see Ref.~\cite{DGH95}). 

\vspace{-0.003cm}
\section{POLYAKOV LOOP MODELS IN PURE GAUGE SU(2)}

Mean-field theory, dimensional analysis, Z(2) symmetry, reality of $l_1$ 
in SU(2), power expansion in $l_1^2$ imply the following simple form for
the effective free energy:
\begin{equation}
{\cal V} = \left(-\frac{b_2}{2} \, l_1^2 + 
\frac{b_4}{4} \, l_1^4\right)\, T^4 \;, \;\;\;b_2>0\;, \; b_4>0\;.
\end{equation}
The applicability domain of this model (called model A in the following)
should be a region above $T_c$, but not so close to $T_c$ that mean-field 
is spoiled, in which $l_1$ is small enough to make $l_1^6$, $l_1^8$, ...  
terms negligible. The minimum of ${\cal V}$ is obtained for $l_1^2=b_2/b_4$ 
and leads to 
\begin{equation}
\frac{p}{T^4}=-\frac{{\cal V}_{\mbox{\footnotesize min}}}{T^4}
 = \frac{b_4}{4}\,l_1^4\;.
\end{equation}

\begin{figure}[htb]
\includegraphics*[scale=0.4]{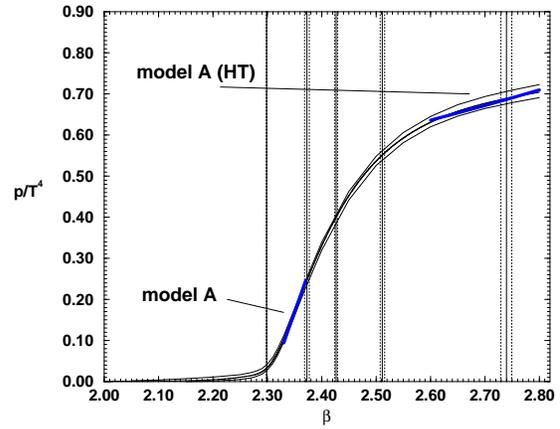}
\vspace{-1.5cm}\caption{Comparison of the model A, for both low and high
temperature regimes, with lattice data for the pressure.}
\label{mod_a}\vspace{-0.7cm}
\end{figure}

According to this model, for {\em constant} $b_4$, $p/T^4$ should go linear
with $l_1^4$. We find that the function $(b_4/4)\,l_1^4$ fits the lattice 
data for the pressure in the region $2.33 \leq 
\beta \leq 2.37$, i.e. $1.11 \,T_c \,\sleq\, T \,\sleq\, 1.27 \,T_c$, 
with $b_4= 261.1(6.7)$ and $\chi^2$/(d.o.f.)=0.79 (see Fig.~\ref{mod_a}).

For high temperatures, one could expand the effective free energy
in powers of $(1-l_1^2)$, thus getting
\begin{equation}
{\cal V}_{\mbox{\footnotesize HT}} = \left(C-\frac{b_2}{2} \, l_1^2
+\frac{b_4}{4} \, l_1^4\right)\, T^4 \;, 
\end{equation}
which leads to 
\begin{equation}
\frac{p}{T^4} = -\frac{{\cal V}_{\mbox{\footnotesize HT,min}}}{T^4}
= C+\frac{b_4}{4}\,l_1^4\;.
\end{equation}
There is compatibility of this functional form with lattice data for constant 
values of $C$ and $b_4$ in the region $2.60 \leq \beta \leq 2.80$, i.e. 
$2.63 \,T_c \,\sleq\, T \,\sleq\, 4.80 \,T_c$, with $b_4 = 19.7(4.8)$, 
$C=0.547(31)$ and $\chi^2$/(d.o.f.)=0.18 (see Fig.~\ref{mod_a}).

As a first variant of the model A, we consider the inclusion of 
the $l_1^6$ term in the effective free energy (model B):
\begin{equation}
{\cal V} = \left(-\frac{b_2}{2} \, l_1^2 + \frac{b_4}{4} 
\, l_1^4 + \frac{b_6}{6} \, l_1^6\right)\, T^4 \;,
\end{equation}
which leads to 
\begin{equation}
\frac{p}{T^4} = -\frac{{\cal V}_{\mbox{\footnotesize min}}}{T^4}
= \frac{b_6}{3}\,l_1^6 + \frac{b_4}{4}\,l_1^4 \;.
\end{equation}

We find compatibility with the lattice data for the pressure over a wider 
region than in the case of model A, more precisely in the range 
$2.32 \leq \beta \leq 2.70$, i.e. $1.07 \,T_c \,\sleq\, T \,\sleq\, 3.56 
\,T_c$, with $b_4= 350.3(6.1)$, $b_6=-1158(33)$ and $\chi^2$/(d.o.f.)=0.73. 
A negative value for $b_6$ would be problematic if 
the absolute value of $l_1$ would be allowed to become arbitrarily large, 
which is not the case here.
For high temperatures, using $(1-l_1^2)$ as expansion parameter we get
\begin{equation}
\frac{p}{T^4} = -\frac{{\cal V}_{\mbox{\footnotesize HT,min}}}{T^4}
= C+\frac{b_6}{3}\,l_1^6 + \frac{b_4}{4}\,l_1^4 \;,
\end{equation}
which agrees with lattice data for the pressure in the region 
$2.50 \leq \beta \leq 2.80$, i.e. $1.93 \,T_c \,\sleq\, T \,\sleq\, 
4.80 \,T_c$, with $b_4= 162(39)$, $b_6=-447(134)$, \- $C=0.258(65)$ 
and $\chi^2$/(d.o.f.)=0.15.

\begin{figure}[htb]
\vspace{-0.5cm}
\includegraphics*[scale=0.4]{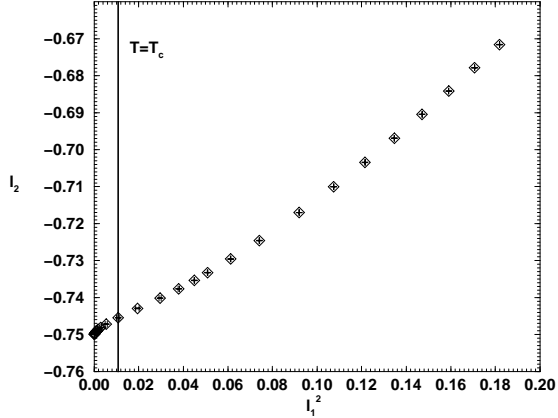}
\vspace{-1.5cm}\caption[]{$l_2$ {\it vs} $l_1^2$ on a $16^3\times 4$ lattice. 
There is a roughly linear dependence in two regimes: for $1.11 \,T_c \,\sleq\,
T \,\sleq\, 1.27 \,T_c$ (corresponding to $0.038 \,\sleq\, l_1^2 \,\sleq\, 
0.061$) and for $2.63 \,T_c \,\sleq\, T \,\sleq\, 4.80 \,T_c$ (corresponding 
to $0.135 \,\sleq\, l_1^2 \,\sleq\, 0.182$).}
\label{l2_l1}\vspace{-0.7cm}
\end{figure}

Finally, we consider the model C obtained by model A with
the inclusion of terms with the charge-2 Polyakov loop $l_2$: 
\begin{equation}
\frac{{\cal V}}{T^4}\! =\! \left(\!-\frac{b_2}{2} \, l_1^2 + \frac{b_4}{4} 
\, l_1^4 + h l_2+\frac{a_2}{2} \, l_2^2 + \xi\, l_1^2\, l_2 \! \right)\,,
\end{equation}
leading to 
\begin{equation}
\frac{p}{T^4} = -\frac{{\cal V}_{\mbox{\footnotesize min}}}{T^4}
= \frac{b_4}{4}\,l_1^4 -h l_2-\frac{a_2}{2}\, l_2^2 
\end{equation}
and
\begin{equation}
l_2=-\frac{h+\xi \, l_1^2}{a_2}
\end{equation}
For the high temperature version of this model, the only difference 
is an additive constant in the r.h.s. of the expression for $p/T^4$.
The comparison with lattice data shows that the inclusion of the terms 
with $l_2$ does not improve drastically the quality of the fit in
comparison with the model A.
On the other side, the linear dependence of $l_2$ with $l_1^2$ is roughly 
satisfied ($\chi^2$/(d.o.f.)\sleq 2 (see Fig.~\ref{l2_l1}) in both regions 
where the model A works, i.e. for $1.11 \,T_c \,\sleq\, T \,\sleq\, 1.27 
\,T_c$ and for $2.63 \,T_c \,\sleq\, T \,\sleq\, 4.80 \,T_c$. This 
indicates that the behavior of $l_2$ in $T$ above $T_c$ is driven by the 
Polyakov loop $l_1$. The relatively large $\chi^2$ can be explained by the 
very small error bars both in $l_1$ and in $l_2$ which make non-negligible 
higher powers of $l_1$ and $l_2$ in the effective model.

\section{CONCLUSIONS}

Lattice data show that $p/T^4$ has a roughly linear behavior in a 
region centered around $1.2 \,T_c$ and in a region centered around 
$3.5 \,T_c$; in these regions also $l_1^4$ exhibits a linear behavior, while
$l_2$ behaves linearly with $l_1^2$. We have shown that both these evidences 
are in accord with simple mean-field effective models of the Polyakov loop.

\end{document}